\documentclass[aps,prd,amsart,amsmath,twocolumn,showpacs]{revtex4}

\usepackage{tikz,xcolor}
\usepackage[colorlinks = true,
linkcolor = blue,
urlcolor  = blue,
citecolor = blue,
anchorcolor = blue]{hyperref}
\definecolor{lime}{HTML}{A6CE39}
\DeclareRobustCommand{\orcidicon}{%
	\begin{tikzpicture}
		\draw[lime, fill=lime] (0,0)
		circle [radius=0.16]
		node[white] {{\fontfamily{qag}\selectfont \tiny ID}};
		\draw[white, fill=white] (-0.0625,0.095)
		circle [radius=0.007];
	\end{tikzpicture}
	\hspace{-2mm}
}

\foreach \x in {A, ..., Z}{%
	\expandafter\xdef\csname orcid\x\endcsname{\noexpand\href{https://orcid.org/\csname orcidauthor\x\endcsname}{\noexpand\orcidicon}}
}

\usepackage[T1]{fontenc}
\usepackage[latin9]{inputenc}
\usepackage{color}
\usepackage{array}
\usepackage{amstext}
\usepackage{graphicx}
\usepackage{esint}
\usepackage{rotating}
\usepackage{appendix}
\usepackage{epsfig}
\usepackage{graphics}
\usepackage{latexsym}
\usepackage{amssymb}
\usepackage{rotating}
\usepackage{subfigure}
\usepackage{bm}
\usepackage{amsmath,amssymb}
\usepackage{blindtext}

\begin{document}


\title{Improved determination of strange distribution function from the global analysis using BHPS model}

\author{Maral Salajegheh$^{a}$\orcidA{}}
\email{m.salajegheh@stu.yazd.ac.ir}

\author{S. Mohammad Moosavi Nejad$^{a,b}$\orcidB{}}
\email{mmoosavi@yazd.ac.ir }

\author{S. Atashbar Tehrani$^{b}$\orcidC{}}
\email{atashbar@ipm.ir
	}

\affiliation {
$^{a}$Physics Department, Yazd University, P.O.Box 89195-741, Yazd, Iran   \\
$^{b}$School of Particles and Accelerators, Institute for Research in Fundamental Sciences (IPM), P.O.Box 19395-5531, Tehran, Iran 
 }

\date{\today}

\begin{abstract}
We study the impact of intrinsic strange (IS) component 
of nucleon sea on the global analysis of parton distribution functions (PDFs) considering a wide range of experimental data.
To this aim, we consider two scenarios on the basis of BHPS model results for the IS distribution. In the first scenario, we apply the results presented through the BHPS model and in the second scenario we use its evolved distributions. For each scenarios, we present the limit of 
the IS probability $ {\cal P}_5^{s\bar{s}} $ for the standard tolerance criteria $ \Delta\chi^2=1 $ and 
$ 18.112 $ at $ 1\sigma $ and $ 4\sigma $ levels.
Our results show that the experimental data can tolerate an IS component 
with a greater probability $ {\cal P}_5^{s\bar{s}} $ if one employs the second scenario. We obtain $ {\cal P}_5^{s\bar{s}}\approx 0.01 $ and 
$ {\cal P}_5^{s\bar{s}}\approx 0.025 $ for $ \Delta\chi^2=1 $ and $18.112$, respectively, at the $ 4\sigma $ level.
We also calculate the ratio of strange-to-light sea-quark densities $r_s$ in the proton both including and excluding the IS component. 
Our results show that one can obtain a higher value for the ratio $r_s$ if the IS component is included.


\end{abstract}


\maketitle

\section{Introduction}\label{sec:one} 

It is well-known that the global analysis of experimental data
is the best and most reliable way to extract the nucleon parton distribution functions (PDFs) \cite{Harland-Lang:2014zoa,Jimenez-Delgado:2014twa,Abramowicz:2015mha,Dulat:2015mca,Alekhin:2017kpj,MoosaviNejad:2016ebo} as well as the polarized PDFs~\cite{
Jimenez-Delgado:2014xza,Sato:2016tuz,Shahri:2016uzl,Khanpour:2017cha}, the nuclear modifications of PDFs~\cite{  deFlorian:2011fp,Kovarik:2015cma,Eskola:2016oht,Klasen:2017kwb}, and the fragmentation functions of partons~\cite{Nejad:2015fdh, MoosaviNejad:2020svj,Delpasand:2020mqv,Nejad:2016epx,Delpasand:2019xpk,MoosaviNejad:2020vqc,MoosaviNejad:2017rvi,Soleymaninia:2017xhc,MoosaviNejad:2017bda,Nejad:2015far,MoosaviNejad:2016scq}.
Actually, these quantities correspond to the nonperturbative aspects of nature and  experimental data will remain, at least for the next few years, the main source of information on them. 
Although, in some theoretical works on basis of lattice QCD calculations and continuum QCD studies  these quantities have been computed  theoretically with large uncertainties, see for example  Refs.~\cite{ Lin:2020rut,Ding:2019qlr}.\\
In the case of PDFs, much progress has been achieved in past few decades. 
However, there are still some different issues which must  be studied 
to make more precise predictions for hadronic collisions, specifically, at the CERN Large Hadron Collider (LHC) and  the Relativistic Heavy Ion Collider (RHIC). One of the important issues is
the accurate determination of the sea quark distributions which we intend to study in this work. This would be possible using the xFitter program~\cite{Alekhin:2014irh} and its facilities. It is now possible  to  study the PDFs in a faster way  in some global analyses including a wide range of  experimental data
at an arbitrary perturbative approximation up to  next-to-next-to-leading order (NNLO) (see for example Refs~\cite{  Bertone:2016ywq,Giuli:2017oii}).

The sea quark distributions extracted from the  global analysis of PDFs
are usually referred to as {\it extrinsic} sea quarks in the scientific literature. In fact, these kinds of sea quarks,  specifically the heavy quark (c,b) sea, arise from the perturbative QCD mechanism and the splitting of gluons into the quark-antiquark pairs in the 
Dokshitzer-Gribove-Lipatov-Alterelli-Parisi (DGLAP) evolution equations~\cite{Altarelli:1977zs,Gribov:1972ri}. 
Although, it is now well-established 
that other kinds of sea quarks can exist inside the nucleon which are known as the {\it intrinsic} ones.
Historically, the existence of these intrinsic sea quarks in the proton was first suggested by Brodsky, Hoyer, Peterson, and Sakai (BHPS) in 1980~\cite{Brodsky:1980pb,Brodsky:1981se}.
In comparison with the extrinsic sea quarks, the intrinsic ones arise from  the nonperturbative mechanism in origin,
and their contributions to the nucleon wave function become important and considerable at relatively large values of the Bjorken scaling variable $x$. \\
The first phenomenological attempt to study the impact of an intrinsic quark (specifically, the charm) component of the nucleon sea on the global analysis of PDFs was made by Harris \textit{et al.}~\cite{Harris:1995jx}, so their analysis suggests a probability of $ 0.86\pm0.6 $\% for the intrinsic charm (IC). 
Their result  was smaller than the BHPS theoretical estimation~\cite{Brodsky:1980pb,Brodsky:1981se} which was obtained (about)  $ 1 $\%  using the diffractive production of $ \Lambda_c $.
The first comprehensive global analysis of PDFs including the IC component was made by the CTEQ Collaboration~\cite{Pumplin:2007wg,Nadolsky:2008zw} using a wide range of the hard scattering data. They obtained a limit up to three times larger for the IC probability in comparison with the BHPS estimation. 
The studies of CTEQ Collaboration were followed in  Ref.~\cite{Dulat:2013hea} where the authors reported a larger value for the IC probability.
In the following, Jimenez-Delgado~\textit{et al.}\cite{Jimenez-Delgado:2014zga} presented a new global
analysis of PDFs considering the IC component. Their conclusion was that
the IC probability can not be greater than $0.5\%$, although it was criticized by Brodsky and Gardner~\cite{Brodsky:2015uwa}. 
Very recently, the authors of Ref.~\cite{Hou:2017khm} have studied
the possibility of a (sizable) nonperturbative contribution to the charm PDF in the context of the CTEQ-TEA global QCD analysis.
They found that the  momentum fraction of charm quark must be less than (about) 2\% for the BHPS IC model.
Consequently, there is  still no agreement concerning the allowed amounts of IC probability in the nucleon sea.

In spite of the global analysis of PDFs for studying the impact of nucleon intrinsic charm component, there are no  similar global  analysis for the intrinsic light quarks. Although, despite some studies on the distributions of the intrinsic light quarks in the nucleon and their related probabilities \cite{Chang:2011vx,Chang:2011du,Chang:2014lea,Salajegheh:2015xoa,An:2017flb}, a simultaneous study of the intrinsic and extrinsic light quarks has not been performed yet in any global analysis of a wide range of the experimental data. Among the sea quark distributions of the nucleon, the strange one has an important role in better understanding the nucleon structure. It would be also useful for describing some DIS processes at the LHC. Moreover, the study of $s-\bar{s}$ asymmetry has been always very interesting~\cite{Salajegheh:2015xoa,Vega:2015hti} because
both experimental and theoretical analysis indicate its existence in the nucleon sea. Therefore, apart from the IC component, it seems that a simultaneous study of the intrinsic strange (IS) component and  the extrinsic one  is motivated in a global analysis of PDFs and would be of particular interest for researchers. \\
In this work, we will study this simultaneous analysis by virtue of the xFitter program~\cite{Alekhin:2014irh} and by considering a wide range of the experimental data.
For this purpose, considering two scenarios we use the result of the BHPS model for the IS distribution in the proton. 
In the first scenario, we apply the results presented through the BHPS model, and in the second one  we use its evolved distribution. For each scenarios, we present the limit of 
the IS probability $ {\cal P}_5^{s\bar{s}} $ for the standard tolerance criteria $ \Delta\chi^2=1 $ and 
$ 18.112 $, and also at $ 1\sigma $ and $ 4\sigma $ level.

The contents of the present paper are as follows. In Sec.~\ref{sec:two},
we shall first describe the BHPS model briefly, and then introduce its result for the IS
distribution in the proton. We will also discuss the evolution of the IS distribution using
the non-singlet DGLAP equation and present a simple functional form for the evolved 
BHPS IS distribution. In Sec.~\ref{sec:three}, we present a detailed explanation of the QCD analysis such as the treatment of heavy-flavor contributions,
PDF parameterization and the procedure for determining the limit of 
the IS probability. In Sec.~\ref{sec:four1}  we explain our fitting procedure by describing  the essential concepts of the Hessian or error matrix approach and Sec.~\ref{sec:four} contains the global analyses of PDFs. We also determine the limit of the IS probability in this section. In Sec.~\ref{sec:five},
we compare the extrinsic strange distribution in the proton obtained from the 
global QCD analysis with the results of the total strange distribution
obtained via considering an IS component. We also calculate the ratio of strange-to-light sea-quark densities in the proton.
Finally, our results and conclusions are summarized in Sec.~\ref{sec:six}.

%
\section{Intrinsic strange distribution}\label{sec:two}
Unlike the extrinsic sea quark distributions
that should be determined phenomenologically in a global analysis of experimental data, 
the intrinsic ones can be calculated theoretically in a light-cone picture of the nucleon at fixed light-front time.
Actually, these components can arise from the nonperturbative fluctuations of the nucleon state to the five-quark
states $\vert uudq \bar{q} \rangle$ where $q=u, d, s$
(and also heavy quarks $c$ and $b$), or the virtual meson-baryon states in the mesonic cloud model (MCM) framework~\cite{Thomas:1983fh,Signal:1987gz}.
One can find a review of these models in Refs.~\cite{Salajegheh:2015xoa,Pumplin:2005yf,Hobbs:2013bia}.
It is worth noting in this context that, in Ref.~\cite{Pumplin:2005yf} authors have presented
a model in which the light-cone probability distributions are derived directly from Feynman diagrams.
Although, there are some clear evidences for the existence of the intrinsic quarks in proton~\cite{Brodsky:2015fna}, but their experimental presence has not been yet confirmed definitively. Then, the only remaining task is the  determination of their presence probabilities in the nucleon.

According to the BHPS model, if one neglects the effect of the transverse momentum in the five-quark
transition amplitudes, the probability distributions of intrinsic quarks $ q, \bar q $ in the five-quark Fock
state $\vert uudq \bar{q} \rangle$ are given by
\begin{eqnarray}\label{eq1}
P(x_1, \dots, x_5)= {\cal N} \delta \left(1-\sum _{i=1}^5 x_i\right)
\left[M^2-\sum _{i=1}^5 \dfrac{m_i^2}{x_i}\right]^{-2},
\end{eqnarray}
where $ m_i $ and $ x_i $ are the mass and the momentum fraction carried by the quark $ i $, respectively. 
In Eq.~(\ref{eq1}), $ M $  is the mass of parent hadron, proton in our case, and
$ {\cal N} $ is the normalization constant which is determined by the following relation 
\begin{equation}\label{eq2}
{\cal P}_5^{q\bar{q}}= \int_0^1 dx_1 \dots dx_5 P(x_1, \dots, x_5),
\end{equation}
where $ {\cal P}_5^{q\bar{q}} $ is the $\vert uudq \bar{q} \rangle$-state probability within the
proton. For the case of heavy quarks, it is expected to be roughly proportional to the quark mass as $ 1/m_q^2 $. The momentum distribution of the intrinsic quark $ \bar q $ is obtained by integrating Eq.~\eqref{eq1} over $ x_1, \cdots ,x_4 $. \\
For heavy quarks, BHPS assumed that the masses of nucleon and light quarks are
negligible in comparison with the heavy ones and obtained  the $x$-distribution of the IC component 
in proton, analytically. Although, for the case of light quarks one can not consider such simplifying assumption and 
then  the intrinsic light distributions must be calculated numerically, see  Refs.~\cite{Chang:2011vx,Salajegheh:2015xoa}.
In Ref.~\cite{Salajegheh:2015xoa}, the IS distribution at the initial scale $Q_0=0.5$~GeV is determined from the BHPS model and it is fitted to a simple functional
form as
\begin{eqnarray}\label{eq3}
xs(x)_{int}=226.531\,x^{2.449}(1-x)^{8.433}.
\end{eqnarray}
In this way, the momentum fraction carried by the IS component in  five-quark state $\vert uuds \bar{s} \rangle$ with $s=\bar s$, i.e.
\begin{eqnarray}\label{eq4}
\langle x \rangle_{\textrm{IS}}=\int_0^1 2xs(x)_{int}\,dx,
\end{eqnarray}
is equal to $ \langle x \rangle_{\textrm{IS}}=0.41 $ which is smaller than the one carried by the IC in 
five-quark state $\vert uudc \bar{c} \rangle$  ($ \langle x \rangle_{\textrm{IC}}=0.57 $~\cite{Brodsky:1980pb}) at the same energy scale $Q_0=0.5$~GeV.
Note that, by neglecting the probability of the $\vert uudq \bar{q} \rangle$-state in the proton,
the quark number condition $ \int_0^1 xf(x)dx=1 $, ($ f=q, \bar q $) should be always satisfied.

\begin{figure}[t!]
\centering
\includegraphics[width=8.6cm]{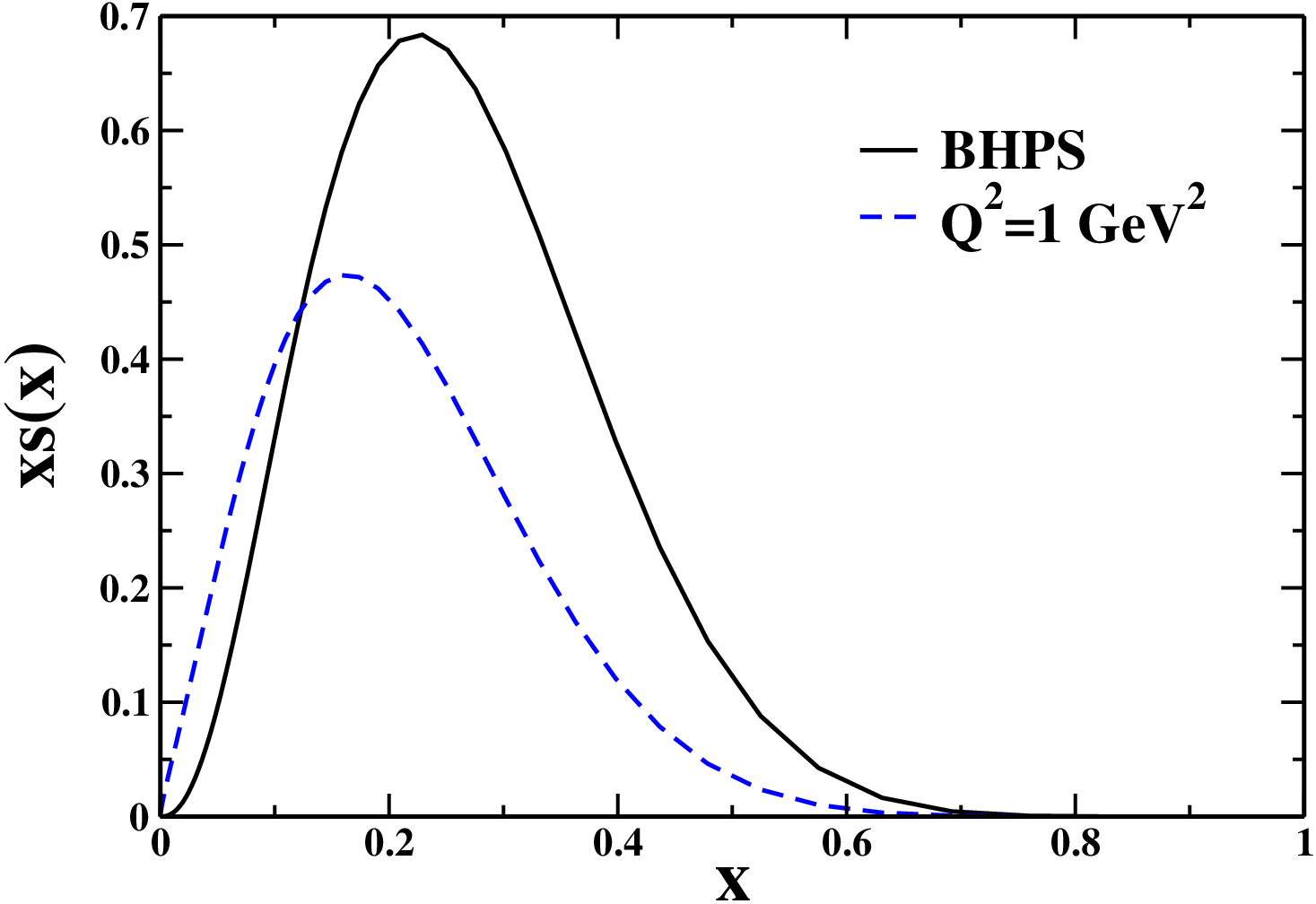}
\caption{Comparison between the BHPS result (\ref{eq3}) for the IS distribution (solid line)
and its evolved distribution (dashed line) at the scale $ Q=1$~GeV (\ref{eq5}).}
\label{fig:fig1}
\end{figure}

One of the main issues concerning the intrinsic quark distributions is their evolution to higher energy scales $Q^2$.
In Ref.~\cite{Lyonnet:2015dca}, it is
shown that the $Q^2$-evolution of the intrinsic quark distributions can be well-controlled by 
non-singlet evolution equations. An interesting advantage of this technique is to evolve the intrinsic quark distributions independent of  the gluon and other PDFs.
On the other hand, in Refs.~\cite{Chang:2011vx,Chang:2011du,Chang:2014lea}
it is shown that if the intrinsic light quark distributions are evolved from the initial scales $Q_0=0.3$ or $0.5$ GeV, a more convenient fit can be achieved.
Therefore, using the non-singlet evolution equations to evolve Eq.~\eqref{eq3} to the higher scale $Q=1$~GeV, one gets
\begin{eqnarray}\label{eq5}
xs(x)_{int}=21.5296\,x^{1.3874}(1-x)^{7.4453}.
\end{eqnarray}
Note that, the momentum fraction carried by this evolved IS distribution is  $\langle x \rangle_{\textrm{IS}}=0.27$
which is smaller than the one obtained at the scale $Q_0=0.5$~GeV, i.e. $ \langle x \rangle_{\textrm{IS}}=0.41 $.\\
Fig.~\ref{fig:fig1} shows the behavior of IS distribution given by  Eqs.~\eqref{eq3} and \eqref{eq5}.
As is seen, the peak position of evolved IS distribution (dashed line)  is shifted towards lower values of $x$ and its amount decreases in the peak region, by as much as $32\%$. Then, it seems that if one uses these different IS distributions in the analysis, various results can be obtained for any physical quantity sensitive to the strange content of the proton.

Now, having the IS distribution from the BHPS model, it is important to estimate its probability ($ {\cal P}_5^{s\bar{s}} $) in the proton. It should be noted that, this quantity has been already
calculated in Refs.~\cite{Chang:2011vx,Chang:2011du,Chang:2014lea} and \cite{An:2017flb}
so that many different results have been reported for its value. 
In Ref.~\cite{Chang:2011vx}, Chang and Peng assumed that the probability of intrinsic light sea quarks (with the mass $m_q$) is proportional to $ 1/m_q^2 $, similar to the heavy quark state. Then, they concluded $ {\cal P}_5^{s\bar{s}}/[ \frac{1}{2}({\cal P}_5^{u\bar{u}}+{\cal P}_5^{d\bar{d}})]=m_{\bar u}^2/m_{\bar s}^2\approx 0.36 $ with $ m_{\bar u}=0.3 $ GeV$ /c^2 $ and $ m_{\bar s}=0.5 $ GeV$ /c^2 $. 
By comparing the results based on the BHPS model and the data for $ \bar d-\bar u $ and $ \bar u+\bar d -s-\bar s $, they have also extracted the probabilities for the $\vert uudu \bar{u} \rangle$- and
$\vert uudd \bar{d} \rangle$-states as $ {\cal P}_5^{u\bar{u}}=0.176 $ and $ {\cal P}_5^{d\bar{d}}=0.294 $, respectively. This implies the value $ {\cal P}_5^{s\bar{s}}=0.08478 $ for the IS probability in the proton. Their studies are followed in Ref.~\cite{Chang:2011du}
where the $ {\cal P}_5^{s\bar{s}} $ is determined using the $ x(s+\bar s) $ data from the HERMES
Collaboration measurement of charged kaon production in a SIDIS reaction~\cite{Airapetian:2008qf}.
They obtained  $ {\cal P}_5^{s\bar{s}}=0.024 $ through the analysis with the initial scale $ Q_0=0.5 $~GeV,
and $ {\cal P}_5^{s\bar{s}}=0.029 $ for the case of $ Q_0=0.3 $~GeV. In a newer version of their studies \cite{Chang:2014lea} they extracted the $ {\cal P}_5^{s\bar{s}} $ in different scenarios which led to the
different values ranging from $ 0.0 $ to $ 0.111 $.
Recently, An and Saghai~\cite{An:2017flb} have analyzed the 
ratios for light  quark-antiquark pairs (i.e. $ u\bar u /d \bar d $ and $ s\bar s /d \bar d $) and found $ {\cal P}_5^{s\bar{s}}=0.057 $. Their analysis was based on the ratios measured from  pion and kaon electroproduction cross
sections reported by
the CLAS Collaboration~\cite{Park:2014zra}. 
In spite of the analyses mentioned above for determination of the probability of  IS distribution in the proton, 
a simultaneous study of the intrinsic strange component and  the extrinsic  sea quarks in a global
analysis of a wide range of the experimental data has not been done, yet.
One  advantage of this study is to obtain the new limits on the intrinsic strange probability in the nucleon, consistent with the large number of the experimental data.

%
\section{Technical detail of QCD analysis}\label{sec:three}
As was already mentioned, in our study we use the open source QCD fit 
framework for PDF determination xFitter~\cite{Alekhin:2014irh}. Meanwhile, the QCDNUM evolution code~\cite{Botje:2010ay} 
is used for the evolution of parton densities. To consider the contributions of heavy flavors,
we apply the Thorne-Roberts scheme~\cite{Thorne:1997ga}.
We also set the charm and bottom quark masses as $ m_c=1.43 $ GeV and $ m_b=4.5 $ GeV.
The analysis is done at the NLO QCD approximation with the strong coupling constant evaluated at NLO in the $\overline{MS}$ scheme  with $n_f=5$ active flavors  adjusted such that $ \alpha_s^{(5)}(M_Z)=0.118 $ for $ M_Z=91.1876 $ GeV.

In our analysis, in addition to the gluon and valence distributions, i.e. $ xg(x) $, $ xu_v(x) $ and  $ xd_v(x) $, we shall also parametrize the anti-quark distributions $ x\bar u $, $ x\bar d $ and $ x\bar s $, separately. This is due to the fact that we use a wide range of the experimental data, including those obtained from the $ W $ and $ Z $ bosons production in hadronic collisions.
We shall also assume that the sea quark and anti-quark distributions are the same.
The PDFs are parametrized at the initial
scale $ Q_0^2=1 $ GeV$^2$ and then evolved to the energy scale of experimental data. 
According to the factorization  theorem of the QCD-improved parton model ~\cite{Collins:1989gx}, by the convolution of these evolved PDFs  with the hard-scattering coefficients one can obtain the theoretical cross sections.

The optimal functional form of the PFDs parameterization is found by the saturation technique of $\chi^2$~\cite{Adloff:2000qk,Aaron:2009aa}.
Following this approach, the parameterizations used to describe the quark and gluon PDFs at the initial scale $Q_0^2=1$~GeV$^2$, read
\begin{eqnarray}\label{eq6}
xg(x) &=& A_g x^{{B_g}}(1-x)^{C_g} - A'_g x^{{B'_g}}(1-x)^{C'_g}, \nonumber \\
xu_{\rm v}(x) &=& A_{u_{\rm v}} x^{B_{u_{\rm v}}}(1-x)^{C_{u_{\rm v}}}(1+E_{u_{\rm v}} x^2), \nonumber \\
xd_{\rm v}(x) &=& A_{d_{\rm v}} x^{B_{d_{\rm v}}}(1-x)^{C_{d_{\rm v}}},\nonumber \\
x \bar u(x) &=& A_{\bar u} x^{B_{\bar u}}(1-x)^{C_{\bar u}}, \nonumber \\
x \bar d(x) &=& A_{\bar d} x^{B_{\bar d}}(1-x)^{C_{\bar d}}, \nonumber \\
x \bar s_{ext}(x) &=& A_{\bar s} x^{B_{\bar s}}(1-x)^{C_{\bar s}}.
\end{eqnarray}
where {\it ext} refers to the extrinsic strange sea distribution. 

In Eq.~\eqref{eq6}, as usual, the parameter $ A_g $ is fixed by the constraint induced by the momentum sum rule.
The parameters $ A_{u_{\rm v}}$ and $A_{d_{\rm v}} $ are determined by the virtue of the quark number sum 
rules for the up and down valence quarks. 
For the gluon parameterization, we also make an additional 
constraint $ C'_g=25 $ as suggested in Ref.~\cite{Abramowicz:2015mha}.
Moreover, to ensure the same behavior of $ x\bar u $ and $ x\bar d $ when $x\rightarrow 0$,
we take some additional constraints as: $A_{\bar{u}}=A_{\bar{d}}$ and $B_{\bar{u}}=B_{\bar{d}}$.
For the strange distribution, we also assume $ A_{\bar{s}}=f_s A_{\bar{d}}/(1-f_s) $ in which
$ f_s $ appears as a free parameter. After imposing these constraints,
the global fit includes $ 16 $ free parameters.\\
It should be also noted that, to avoid the nonperturbative effects, all DIS data included in our global analysis  should satisfy two conditions as $Q^2\geq 3.5$~GeV$^2$
and $W^2\geq 3.5$~GeV$^2$, and all jet production data in hadronic collisions should satisfy $p_T\geq 20$~GeV, as well. 
Then, all remaining PDF parameters are determined by xFitter in minimizing the $\chi^2$-function using the MINUIT
program~\cite{James:1975dr}. 
To estimate the PDF
uncertainties the experimental data uncertainties are propagated to the extracted QCD fit parameters 
using the asymmetric Hessian method~\cite{Pumplin:2001ct} that is applicable in the xFitter framework. The $\chi^2$-function along with the  asymmetric Hessian method are introduced in Section \ref{sec:four1}.  

With the above explanations, the total strange sea distribution in the proton is written as 
\begin{eqnarray}\label{eq7}
xs_{tot}(x)=xs_{fit}(x)+ {\cal P}_5^{s\bar{s}}  xs_{BHPS}(x),
\end{eqnarray}
where, the factor $ {\cal P}_5^{s\bar{s}} $ describes the probability of finding an IS component in the proton. The  term $ xs_{fit}(x)$ is determined in global analysis of the experimental data and for the term $xs_{BHPS}$, one can apply the functional form obtained from the BHPS model \eqref{eq3} or its evolved distribution \eqref{eq5}.

Now, we summarize our procedure for the analysis. At first, we perform a global analysis of PDFs in the framework described in this section without considering any intrinsic component for the strange distribution in the proton. In the following, we repeat our analysis considering the intrinsic strange distribution, as a certain function multiplied by the factor $ {\cal P}_5^{s\bar{s}} $. Then, to construct the total strange
distribution (\ref{eq7}) we add its contribution to the fitted distribution   at the initial scale $Q_0^2=1 $~GeV$^2$. The analysis is repeated several times by varying the 
value of the $ {\cal P}_5^{s\bar{s}} $. 
Note that, during each analysis the value of  $ {\cal P}_5^{s\bar{s}} $ is fixed. In this way, for each data set (and also for all data) we can compute the deviation of 
$ \chi^2 $-values from their central ones, corresponding to
the analysis performed without considering an IS component, see section \ref{sec:four1}. Then, we can investigate the consistency 
of each data set with the IS component and obtain the new limits on the IS probability in the nucleon consistent with the large number of experimental data. For the $ {\cal P}_5^{s\bar{s}} $, we choose a value in the range $ 0.001-0.03$
in our analysis. It should be also noted that for the analysis including an IS distribution,
the momentum sum rule of PDFs  at the scale $Q_0^2$ must be modified as  
\begin{eqnarray}\label{eq8}
\int_0^1 dx\, x \{g(x)+u_v(x)+d_v(x) +2[\bar u(x) \nonumber \\
+\bar d(x) +\bar s_{ext}(x) + {\cal P}_5^{s\bar{s}} \bar s_{int}(x)] \} =1.
\end{eqnarray}
%

\section{ $\chi^2$-minimization approach}
\label{sec:four1}

Here we, briefly, explain our fitting procedure by describing  the essential concepts of the Hessian or error matrix approach  which is a usual minimization method. More details can be found in Refs.~\cite{Pumplin:2001ct,Pumplin:2002vw,Martin:2002aw,Martin:2009iq,Salajegheh:2018hfs,Soleymaninia:2013cxa}.
According to the fit procedure which will be presented in Sec.~\ref{sec:three}, we have 
a set of appropriate parameters for PDFs, i.e., $\{p_i\} (i=1,2,\cdots,n)$,  where $n$ refers to the total number of the fitted parameters.  The basic assumption of the Hessian approach is a quadratic expansion of the $\chi^2_{global}$ around the minimum $\chi^2$ point as
\begin{eqnarray}\label{eq4-1}
\Delta\chi^2_{global}(p)&=&\chi^2_{global}-\chi_{min}^2\nonumber\\
&&\hspace{-0.5cm}=\sum_{i,j}H_{ij}(p_i-p_i^0)(p_j-p_j^0),
\end{eqnarray}
where,  the element of the Hessian matrix $H_{ij}$ is defined as
\begin{eqnarray}\label{eq4-2}
H_{ij}=\left.\frac{1}{2}\frac{\partial^2\chi^2_{global}}{\partial p_i \partial p_j}\right|_{min}.
\end{eqnarray}
Since, the Hessian matrix and its inverse $C\equiv H^{-1}$, which is the error matrix, are symmetric they have a set of n orthogonal eigenvectors $v_{ik}$ with eigenvalues $\lambda_k$: 
\begin{eqnarray}\label{eq4-3}
	&&\sum_{j=1}^{n}C_{ij}v_{jk}=\lambda_k v_{ik},\nonumber\\
	&&\sum_{i=1}^{n}v_{ij}v_{ik}=\delta_{jk}.
\end{eqnarray}
The parameter variation around the global minimum can be expanded in a basis  of rescaled
eigenvectors $e_{ik}=\sqrt{\lambda_k}v_{ik}$ as
\begin{eqnarray}\label{eq4-4}
p_i-p_i^0=\sum_{k=1}^{n}e_{ik}z_k.
\end{eqnarray}
Replacing Eq.~(\ref{eq4-4}) in Eq.~(\ref{eq4-1}) and considering the orthogonality
of $v_k$ we achieve	
\begin{eqnarray}\label{eq4-5}
\Delta \chi^2_{global}=\sum_{k=1}^{n}z_k^2,	
\end{eqnarray}
where $\sum_{k=1}^{n}z_k^2\leq T^2$ is the interior of a sphere of radius T.
The neighborhood parameters are given by
\begin{eqnarray}\label{eq4-7}
p_i(s_k^{\pm})=p_i^0\pm te_{ik}	,
\end{eqnarray}
where $s_k$ is the $k$'th set of PDFs and $t$ is adapted to
make the desired $T=(\Delta \chi^2_{global})^{\frac{1}{2}}$	and $t = T$ in the quadratic approximation.
The PDF uncertainties are estimated using the Hessian matrix as follows:
\begin{eqnarray}\label{eq4-8}
	[\delta f(x)]^2&=&\Delta \chi^2\sum_{i,j=1}^n\frac{\partial f(x,p_i)}{\partial p_i}H_{ij}^{-1}\frac{\partial f(x,p_j)}{\partial p_j},
\end{eqnarray}
where $H_{ij}$ is the Hessian matrix and  $n$ is the number of parameters in the global fit. The $\Delta \chi^2$ value determines the confidence
region and is given 
by
\begin{eqnarray}\label{eq4-9}
	\Delta \chi^2(p)=\sum_{i,j}H_{ij}\delta p_i \delta p_j.
\end{eqnarray}	
The $\Delta \chi^2$	value is calculated considering the confidence
level $P$ which is defined as
\begin{eqnarray}\label{eq4-10}
	P=\int_{0}^{\Delta \chi^2}\frac{1}{2\Gamma\left(\frac{N}{2}\right)}\left(\frac{s}{2}\right)^{\frac{N}{2}-1}exp\left(-\frac{s}{2}\right)ds.
\end{eqnarray}
The numerical value of Eq.~(\ref{eq4-10}) which corresponds to 1$\sigma$
error is $P = 0.6826$. This numerical value is related to a given
number of parameters ($N$) by assuming the normal distribution
in the multiparameter space. In an analysis with  16 parameters, we
achieve $\Delta \chi^2=18.112$. The Hessian method can also be used
to estimate the polarized PDFs and fragmentation functions~\cite{Pumplin:2001ct,Pumplin:2002vw,Martin:2002aw,Martin:2009iq,Salajegheh:2019srg,Salajegheh:2019nea,Salajegheh:2019ach,MoosaviNejad:2018ukp}.
%
\section{fit results}\label{sec:four}
Defining our framework, we are now in a situation to perform our global analysis of PDFs considering an IS component in the proton.  One of the main tasks in this regard  is an appropriate selection of the experimental data. As is well-known, in any global analysis of PDFs, the main constraints on parton densities come from the DIS data which include a wide range of the $x$- and $Q^2$- values. Nevertheless, the DIS data can not  determine the gluon distribution accurately, and can not also specify the sea quark distributions, specifically as the separate flavors.  In fact, for the case of sea quarks the DIS data can just be used to determine  the sum of their distributions in some kinematical regions.
Therefore, to impose further constraints on the gluon and sea quark distributions, various phenomenological groups employ different types of experimental data such as those from the one-jet inclusive productions and also the various measurements of the $W$ and $Z$ boson productions.
Nowadays, with the progresses achieved for the theoretical calculations and having advanced computational tools, many other processes can be also included in the analysis to get  more information over the parton distributions. For example, for the gluon distribution the inclusive isolated prompt photon production can be used in hadronic and nuclear collisions.

In the present work, for the fixed-target DIS data we use the proton structure 
functions $ F_2^p $ determined by BCDMS in scattering of muons over  hydrogen targets~\cite{Benvenuti:1989rh}.
For  DIS data sets, we include the latest high-precision neutral current (NC) and charge current (CC) HERA I+II combined data~\cite{Abramowicz:2015mha} and also the low-$x$ and $Q^2$ measurements from
H1 obtained through the inclusive NC $ e^{\pm}p $ scattering~\cite{Collaboration:2010ry}. These DIS data can constrain the gluon 
and sea quark distributions  in all range of $x$, satisfactorily. In addition,
we use the H1 and ZEUS measurements of the bottom structure function $ F_2^b $~\cite{Aaron:2009af}
and reduced cross section $ \sigma_r^b $~\cite{Abramowicz:2014zub},
and also the HERA combined measurements of the charm production cross section~\cite{Abramowicz:1900rp} that
can be useful to constrain the bottom and charm contents of proton at low values of $x$.
The inclusion of DIS data is completed by considering the one-jet inclusive cross sections
from the H1~\cite{Aaron:2009vs,Aktas:2007aa,Aaron:2010ac} and ZEUS~\cite{Chekanov:2002be,Chekanov:2006xr}, 
and also the dijet cross sections from the ZEUS measurements~\cite{Abramowicz:2010cka}.
For the one-jet inclusive cross sections in hadronic collisions, we use the ATLAS data
at a center-of-mass energies of $7$~TeV and $2.76$~TeV \cite{Aad:2011fc,Aad:2013lpa}
and also the CMS measurements at $7$~TeV \cite{Chatrchyan:2012bja}.
It is now well-established \cite{Rojo:2014kta} that the jet production cross sections reported by the LHC experiments provide a practical tool to achieve the precious information on the gluon and quark PDFs at large values of $x$.
In the case of Drell-Yan differential cross sections in $ pp $ collisions, both high~\cite{Aad:2013iua} and low~\cite{Aad:2014qja}
mass ATLAS measurements at $\sqrt s=7$~TeV are included in our analysis. Finally for the electroweak production of bosons at the Tevatron, we use the CDF data~\cite{Aaltonen:2010zza,Aaltonen:2009ta} 
while for the same measurements at the LHC we use the inclusive $W^\pm$ and
$Z$ rapidity distributions at $7$~TeV from both ATLAS 2010 \cite{Aad:2011dm} and 2011 \cite{Aaboud:2016btc} data sets.
The advantage of these data is to provide more sensitivity to the flavor composition of the sea quark. The same benefit is applicable for the valence-quark distributions at lower-$x$ so that, for instance, the determination
of the strange content of the proton would be possible.

After introducing the experimental data used in our analysis, we present our results obtained through the global fit.
At the first step, we execute a global analysis without including an IS component in the proton.
All experimental data along with the $\chi^2$-values are listed in Table~\ref{table:one}. We have also presented the total $\chi^2$ divided by the number of freedom degrees. i.e. $\chi^2/d.o.f=1.178$.  Note that, the only data which have a rather large $\chi^2$-value are related to the $Z$ rapidity distributions for the central channel (CC) from the ATLAS 2011 data set.   
The optimal values of the PDF parameters \eqref{eq6}  at the initial scale $Q_0^2=1$ GeV$^2$ are listed in Table~\ref{table:two}, so the uncertainties are estimated by the MINUIT program \cite{James:1975dr}.

\begin{table}[t!]
\small
\caption{ Experimental data 
included in our  analysis (see the text for a detailed explanation). 
The values of $ \chi^2 $ and the number of data points $ N_{\textrm{pts}} $  are presented for each data set.}
\label{table:one}

\begin{ruledtabular}
\begin{tabular}{lc}

  Data set & $ \chi^2/N_{pts} $  \\\hline 
  HERA I+II NC~\cite{Abramowicz:2015mha} 																	& 1291.46/1064  \\ 
  HERA I+II CC~\cite{Abramowicz:2015mha} 																	& 114.8/81  \\ 
  H1 $ F_2^b $~\cite{Aaron:2009af}              																	& 3.73/12 \\
  ZEUS $ \sigma_r^b $~\cite{Abramowicz:2014zub}															& 13.07/17 \\
  H1 incl. jets~\cite{Aaron:2009vs,Aktas:2007aa,Aaron:2010ac}											& 45.62/76 \\
  ZEUS incl. jets~\cite{Chekanov:2002be,Chekanov:2006xr,Abramowicz:2010cka}				& 67.79/82 \\
  H1 low $ ep $~\cite{Collaboration:2010ry}																		& 134.98/124 \\
  HERA $ \sigma_r^c $~\cite{Abramowicz:1900rp} 															& 40.85/47  \\ 
  BCDMS $ F_2^p $\cite{Benvenuti:1989rh} 																		& 320.78/328 \\ 
  CDF II $ W $ asym.~\cite{Aaltonen:2009ta} 																	& 21.9/13  \\
  CDF II $ Z $ rap.~\cite{Aaltonen:2010zza} 																	& 29.27/28  \\
  CMS jets ($ 7 $ TeV)~\cite{Chatrchyan:2012bja} 															& 145.78/133  \\
  ATLAS $ W^+ $ 2010\cite{Aad:2011dm} 																	& 15.47/11 \\
  ATLAS $ W^- $ 2010\cite{Aad:2011dm} 																	& 9.58/11 \\
  ATLAS $ Z $ 2010~\cite{Aad:2011dm} 																			& 2.4/8 \\
  ATLAS Drell-Yan~\cite{Aad:2013iua,Aad:2014qja}															& 19.16/21 \\
  ATLAS jets ($ 7 $ TeV)\cite{Aad:2011fc}  																		& 60.35/90  \\ 
  ATLAS jets ($ 2.76 $ TeV)\cite{Aad:2011fc}  																	& 47.14/56  \\ 
  ATLAS $ W^- $ 2011\cite{Aaboud:2016btc} 																& 12.12/11 \\
  ATLAS $ W^+ $ 2011\cite{Aaboud:2016btc} 																& 14.44/11 \\
  ATLAS CC $ Z $ 2011~\cite{Aaboud:2016btc} 																& 46.16/24 \\
  ATLAS high-mass FC $ Z $ 2011~\cite{Aaboud:2016btc} 												 & 4.14/6 \\
  ATLAS peak FC $ Z $ 2011~\cite{Aaboud:2016btc} 														& 12.19/9 \\ \hline
{\bf TOTAL $ \chi^2/d.o.f $} 																							& $ 2650.56/2250=1.178 $ \\

\end{tabular}
\end{ruledtabular}
\end{table}
\begin{table}[t!]
\small
\caption{The optimal values of the input PDF parameters \eqref{eq6} at 
the initial scale $Q_0^2$= 1 GeV$^{2}$ determined by the global analysis of the experimental data reported in Table~\ref{table:one}.}
\label{table:two}

\begin{ruledtabular}
\begin{tabular}{lc}
  
 Parameter   & Best value  \\\hline
    
  $ B_{g} $ 															& $ -0.3698 \pm 0.02833 $  \\ 
  $ C_{g} $ 															& $ 4.5158 \pm 0.2681 $  \\ 
  $ A'_{g}$ 															& $ 1.3973 \pm 0.1138 $  \\ 
  $ B'_{g}$ 															& $ -0.3562 \pm 0.0241 $  \\  
  $ B_{u_{\rm v}} $ 												& $ 0.7923 \pm 0.0082 $  \\ 
  $ C_{u_{\rm v}}$ 												& $ 2.2082 \pm 0.0327 $  \\ 
  $ E_{u_{\rm v}}$ 												& $ -1.0898 \pm 0.0223 $  \\ 
  $ B_{d_{\rm v}}$ 												& $ 0.8932 \pm 0.0215 $  \\ 
  $ C_{d_{\rm v}}$ 												& $ 4.2397 \pm 0.0788$  \\ 
  $ C_{\bar u}$ 														& $ 3.2517 \pm 0.2990 $  \\ 
  $ A_{\bar d}$ 														& $ 0.0909 \pm 0.0052 $ \\ 
  $ B_{\bar d}$ 														& $ -0.1739 \pm 0.0089 $  \\ 
  $ C_{\bar d}$														& $ 10.9148 \pm 1.5679 $  \\ 
  $ B_{\bar{s}} $ 												& $ 0.0644 \pm 0.0610 $  \\ 
  $ C_{\bar{s}} $ 												& $ 4.9168 \pm 0.7289 $  \\ 
  $ f_s $ 																& $ 0.7166 \pm 0.0532 $  \\ 

\end{tabular}
\end{ruledtabular}
\end{table}

Now, having a basic fit to a wide range of experimental data we can investigate, for the first time,
the impact of the IS component of the nucleon on the global analysis as well as the extracted PDFs. 
Therefore, a similar analysis should be done by including the contribution of  IS component so that each analysis is related to a different value of the  $ {\cal P}_5^{s\bar{s}} $.
For example, we can start with a value of ${\cal P}_5^{s\bar{s}}=0.001$ and go to the ${\cal P}_5^{s\bar{s}}=0.03$
by adding a value of $0.001$ in each stage. In the next step, for each data set and also for all data, we must calculate the deviation of 
$\chi^2$-values from their central values corresponding to
the basic analysis listed in Table~\ref{table:one}. Therefore, the sensitivity of each data set (and also the overall analysis) to the IS component of the proton can be investigated.
Moreover, we can obtain the new limits on the IS probability in the nucleon which is consistent with 
a large number of the experimental data. To consider the IS component we apply two scenarios: the first is based on the IS distribution resulting from the  BHPS model \eqref{eq3},  and the second one is related to its evolved distribution \eqref{eq5}. 
 
\begin{figure}[t!]
\centering
\includegraphics[width=8.6cm]{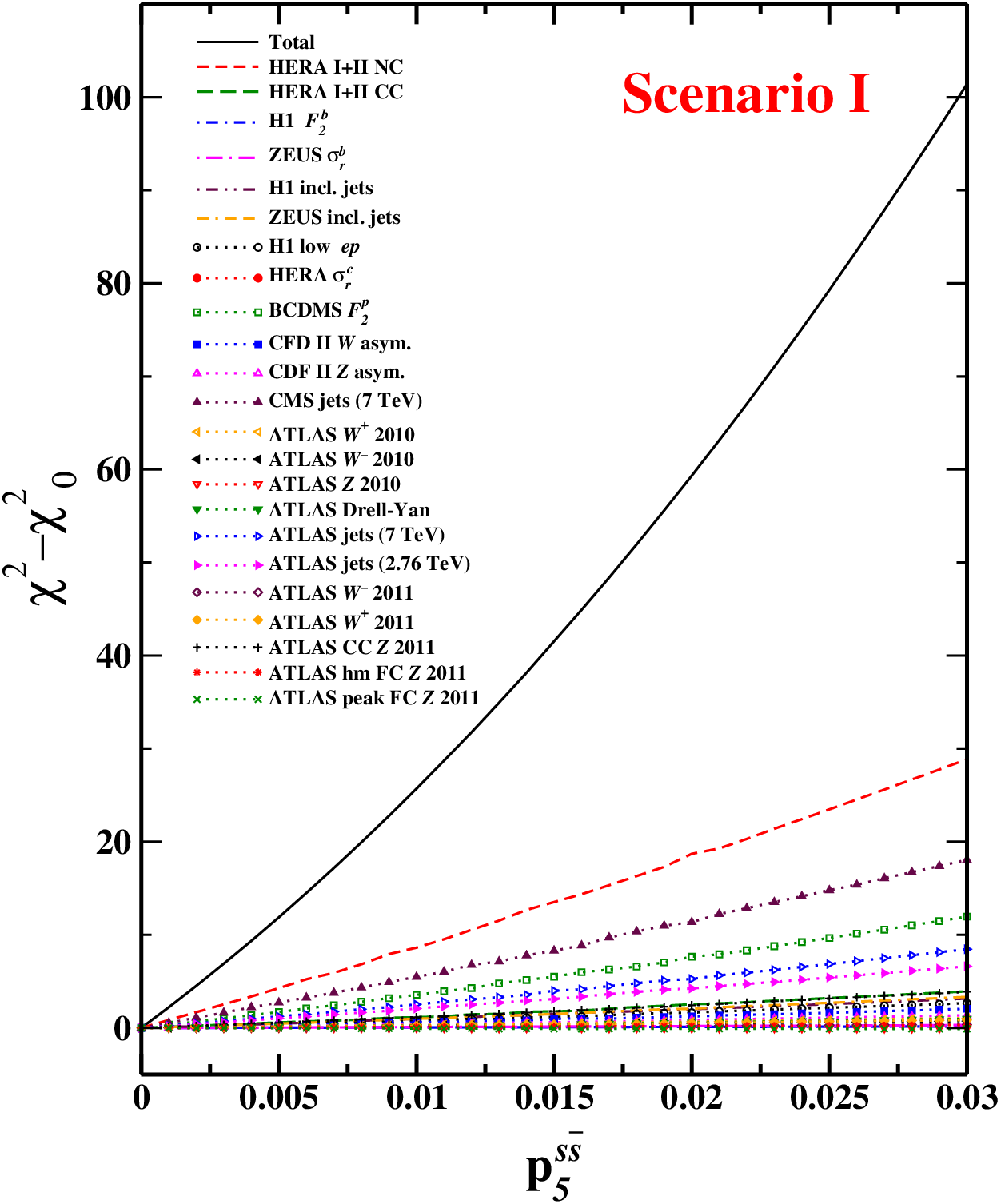}
\caption{The $\chi^2-\chi_0^2$ as a function of the IS probability $ {\cal P}_5^{s\bar{s}} $ in the first scenario. The largest deviation for the $\chi^2$ arises from the NC HERA I+II 
	combined data~\cite{Abramowicz:2015mha} (dashed line),
while the smaller contributions arise from the CMS inclusive jets \cite{Chatrchyan:2012bja}
and BCDMS proton structure functions $ F_2^p $ \cite{Benvenuti:1989rh}.}
\label{fig:fig2}
\end{figure}

Fig.~\ref{fig:fig2} shows the results of our global analysis determined through the first scenario  where we have used Eq.~\eqref{eq3} for the IS distribution of proton.  In this figure, for each data set we have shown the $(\chi^2-\chi_0^2)$-values 
as a function of IS probability ${\cal P}_5^{s\bar{s}}$.  We have also shown the $(\chi^2-\chi_0^2)$-values  for all data (solid line), where the $\chi_0^2$ is obtained through our basic analysis in which the IS component is ignored. Note that, the $\chi^2$ stands for the analysis including the IS component.
Generally, most of data sets have either a little sensitivity or no sensitivity on the IS component  so that their $ \chi^2 $-value does not have a considerable deviation from the corresponding value obtained through the basic analysis. The largest deviation for the $\chi^2$ arises from the NC HERA I+II 
combined data~\cite{Abramowicz:2015mha} (dashed line), while the smaller ones come from the CMS inclusive jets \cite{Chatrchyan:2012bja}
and BCDMS proton structure functions $F_2^p$ \cite{Benvenuti:1989rh}.
An interesting and considerable point is that, when an IS component is considered in the proton  the $Z$ rapidity distributions at $7$~TeV from
the ATLAS 2010 \cite{Aad:2011dm} and 2011 \cite{Aaboud:2016btc} data sets
(for the forward channel (FC) in the $Z$-peak regions) are just experimental data which show a reduction in the $\chi^2$.

In our  QCD analysis of PDFs, we used the Hessian  approach introduced in section \ref{sec:four1}  to estimate the PDFs uncertainties. Following this method, a confidence region is identified by providing the tolerance criterion $ \Delta\chi^2 $.
No unique criterion is defined to select a correct value of $\Delta\chi^2$  in  various global analysis which
have been done yet. In some analysis, a value $\Delta\chi^2=1$ is chosen as a standard tolerance criterion \cite{Alekhin:2017kpj,Jimenez-Delgado:2014twa,Jimenez-Delgado:2014zga}. To be more precise, in such analysis the $\Delta\chi^2$-value is calculated such that the confidence level $P$ becomes the one-$\sigma$-error range
($P = 0.6826$) for only one parameter. In fact, the main argument for this kind of choice is that for the Gaussian distributions
the parameter errors in $\chi^2$-fits should be determined by $\Delta\chi^2=1$ irrespective 
of the number of parameters in the fit \cite{Jimenez-Delgado:2015tma}. This approach is usually called as Hessian methodology, but with no tolerance. In contrary, in some analysis the value of $\Delta\chi^2>1$ is assumed for the tolerance criterion \cite{deFlorian:2011fp,Pumplin:2007wg,Dulat:2013hea} in order for inflating the uncertainty
and counting the tensions among different data sets used in the fit. However, from another point of view, the actual value of $\Delta\chi^2$ depends on the number of parameters 
which are specified in the fit \cite{Brodsky:2015uwa}. Indeed, a value different of $\Delta\chi^2=1$ should be 
assigned for the freedom  degrees $N$. 
For example, if there are $ 16 $ free parameters in the analysis, like ours, the $\Delta\chi^2$-value is assumed as $\Delta\chi^2=18.112$ at the $ 1\sigma$ level. 
This approach is applied in some analysis such as \cite{Kovarik:2015cma,Soleymaninia:2013cxa,Nejad:2015fdh}. 

\begin{figure}[t!]
\centering
\includegraphics[width=8.6cm]{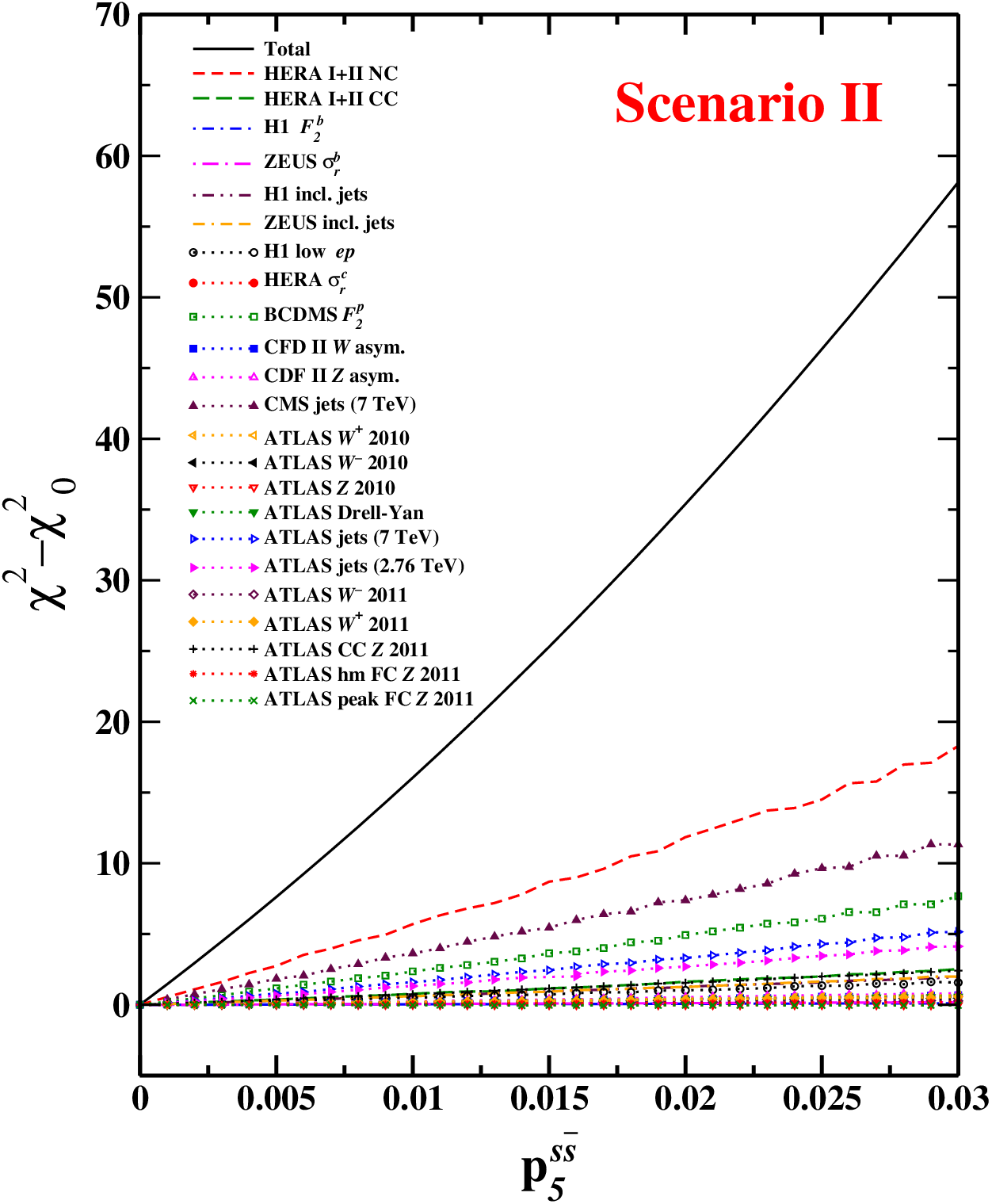}
\caption{As in Fig.~\ref{fig:fig2}, but for the second scenario where we used the evolved BHPS IS distribution  \eqref{eq5}.}
\label{fig:fig3}
\end{figure}

In our analysis, we specify the limit of IS probability ${\cal P}_5^{s\bar{s}}$ both for $\Delta\chi^2=1$ and $18.112$, and also at the $1\sigma$ and $4\sigma$ levels.
Focusing on Fig.~\ref{fig:fig2}, we find that with $ \Delta\chi^2=1 $ at the $ 1\sigma $ level, the experimental
data do not tolerate any IS component in the proton, since even for the ${\cal P}_5^{s\bar{s}}=0.001$ one has $\Delta\chi^2>1$. Nevertheless, at the $4\sigma$ level an IS component 
with the probability ${\cal P}_5^{s\bar{s}}\approx 0.007$ can be considered. Similarly, 
if we consider $\Delta\chi^2=18.112$ corresponding to $ 16 $ free parameters 
at the $1\sigma$ level, we find ${\cal P}_5^{s\bar{s}}\approx 0.008$ 
while the corresponding value  at the $4\sigma $ level  is ${\cal P}_5^{s\bar{s}}\approx 0.016$.
Remember that, the mentioned results are related to the first scenario  where we used the IS distribution \eqref{eq3}. 
However, if we apply the second scenario using Eq.~\eqref{eq5} with a smaller momentum fraction, the results are changed to a considerable extent, see Fig.~\ref{fig:fig3}. As is seen, the sensitivity of  experimental data
to the IS component is decreased in comparison with the previous one, as was expected. In this case, considering $\Delta\chi^2=1$ we
get ${\cal P}_5^{s\bar{s}}\approx 0.001$ and ${\cal P}_5^{s\bar{s}}\approx 0.01$ at the $1\sigma$ and $4\sigma$ levels, respectively.
The corresponding value for the $\Delta\chi^2=18.112$ is ${\cal P}_5^{s\bar{s}}\approx 0.011$ at the $1\sigma$, and
${\cal P}_5^{s\bar{s}}\approx 0.025$ at the $4\sigma$ level. Generally, we can assert that if one uses the evolved BHPS result for the  IS distribution, 
the experimental data can tolerate an IS component 
with a greater probability. Indeed, this conclusion confirms 
the Chang and Peng suggestion \cite{Chang:2011vx,Chang:2011du,Chang:2014lea} so that if the intrinsic light quark distributions are evolved from a lower initial
scale (such as $Q_0=0.5$~GeV or $0.3$~GeV), a more suitable fit can be achieved.\\
As a last point, note that the results of our global analysis  are
accessible in the LHAPDF6 format~\cite{Buckley:2014ana} (For more detail see the Appendix).
These PDF sets can be applied in future studies on the impact of IS distributions on the physical observables.

%
\section{strange content of proton}\label{sec:five} 
\begin{figure}[t!]
\centering
\includegraphics[width=8.6cm]{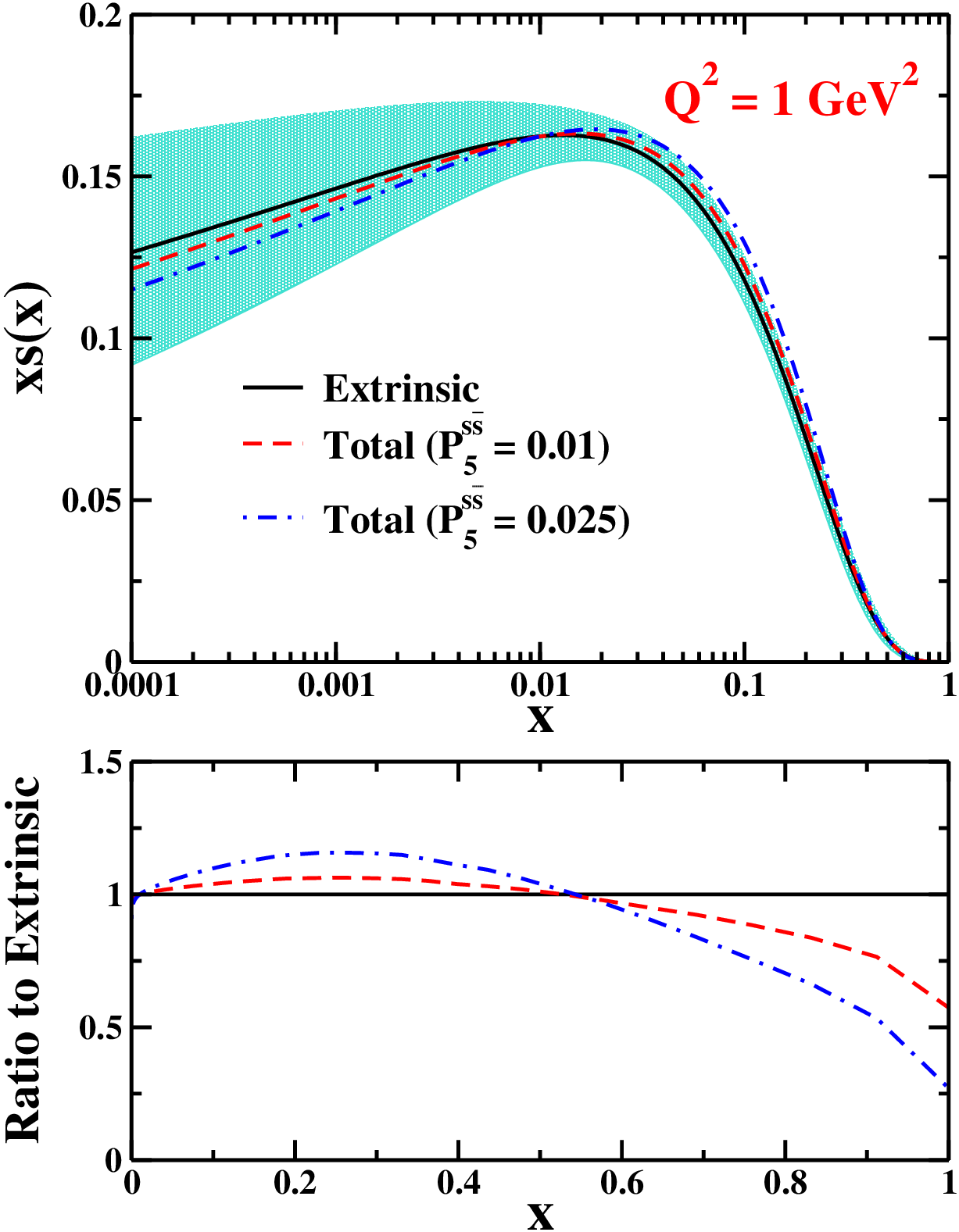}
\caption{Up panel: comparison between the extrinsic strange distribution  in the proton extracted through BHPS model (solid line)
and the total strange distributions including an IS component with the probabilities ${\cal P}_5^{s\bar{s}}=0.01$ (dashed line)
and ${\cal P}_5^{s\bar{s}}=0.025$ (dot-dashed line) at the initial scale $Q_0^2=1$~GeV$^2$.
Down panel:  the total strange distributions normalized to the extrinsic  one.}
\label{fig:fig4}
\end{figure}

As a conclusion of previous section, if one uses an evolved BHPS IS distribution in the proton a greater probability ${\cal P}_5^{s\bar{s}}$ can be obtained, consistent with a wide range of the experimental data. 
In fact, considering the standard tolerance criterion $\Delta\chi^2=1$ we obtained the limit of IS probability as
${\cal P}_5^{s\bar{s}}\approx 0.01$ at the $4\sigma$ level. Moreover, we found that at the same $\sigma$ level 
the limit of IS probability for a greater value $\Delta\chi^2=18.112$ (corresponding to $ 16 $ free parameters),
can even reach the ${\cal P}_5^{s\bar{s}}\approx 0.025$. In this section, considering these two values for $ {\cal P}_5^{s\bar{s}} $,
we study  the strange content of the proton in detail.
To further investigation, the comparisons are made at different values of $Q^2$. \\
Fig.~\ref{fig:fig4} (top panel) shows a comparison between the extrinsic strange distribution extracted through the BHPS model  in the absence of an IS component (solid line), and the total strange distributions in the presence of an IS component with the probabilities ${\cal P}_5^{s\bar{s}}=0.01$ (dashed line)
and ${\cal P}_5^{s\bar{s}}=0.025$ (dot-dashed line) corresponding to two scenarios. For a more quantitative interpretation of Fig.~\ref{fig:fig4}, in the lower panel we showed the ratios of both results  to the extrinsic strange distribution. We have also showed the uncertainty of the extrinsic strange distribution in Fig.~\ref{fig:fig4}.
Note that, the total strange distributions including an IS component with the probability ${\cal P}_5^{s\bar{s}}=0.025$ is not within the error band for large values of $x$, where the contribution of IS component is dominant.
As can be seen, inclusion of an IS component with  the probability of $1\%$ does not change  the strange content of the proton, significantly.
Although, an IS component with the probability of $2.5\%$ can have a remarkable role at the medium- and large-$x$ regions.
The enhancement shown in Fig.~\ref{fig:fig4} is obviously due to the inclusion of an IS component which increases the momentum fraction carried by the strange sea and, in conclusion, the  contribution of strange to the sea quark distribution increases, see Fig.~\ref{fig:fig1}.
The reduction of total strange distributions  for the region $x\gtrsim 0.5$ can be related to the effects of momentum sum rule (\ref{eq8}) for which any reduction (enhancement) in the
momentum fraction carried by a specific parton leads to an enhancement (reduction) in the momentum fraction carried by another
parton.

\begin{figure}[t!]
\centering
\includegraphics[width=8.6cm]{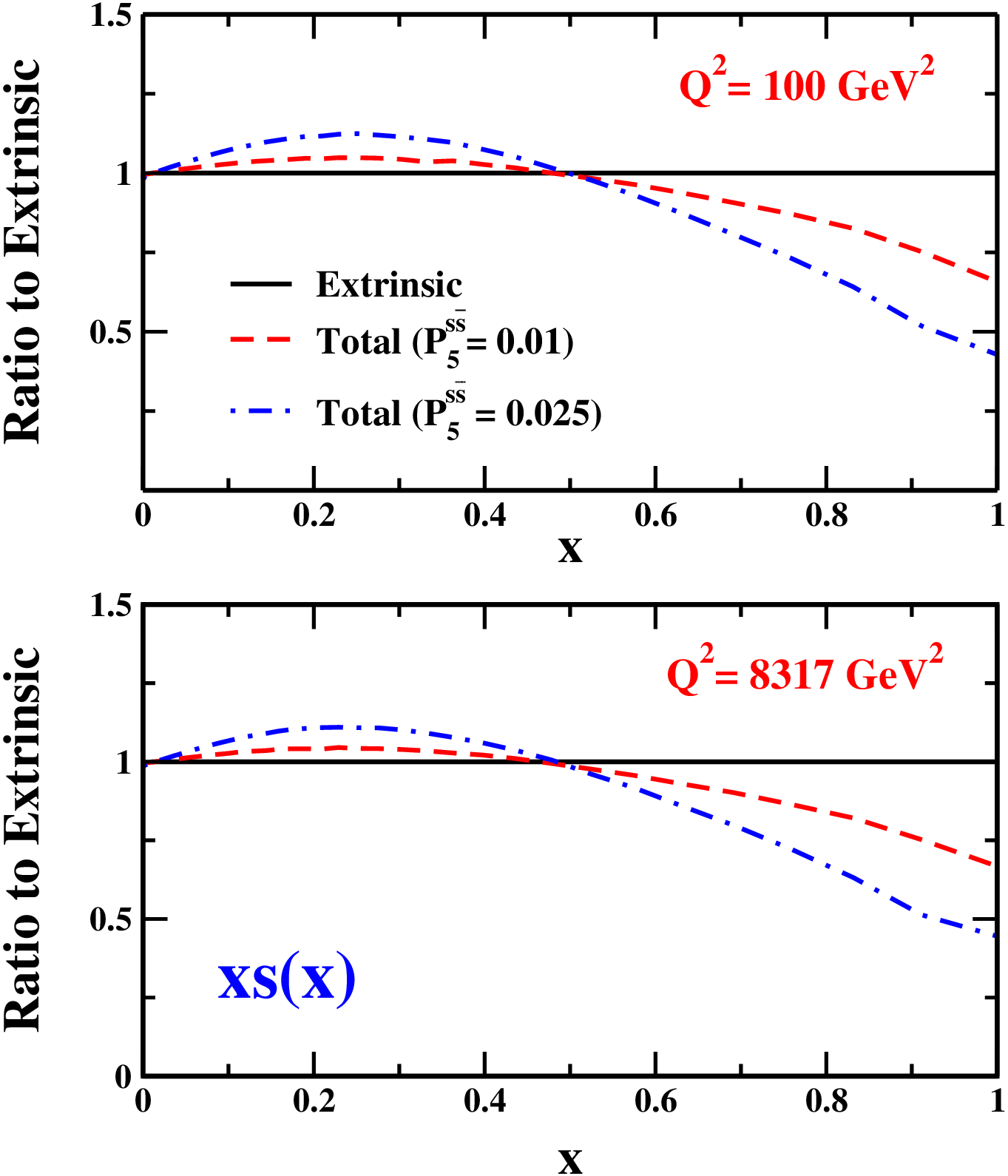}
\caption{ As in Fig.~\ref{fig:fig4}, but for higher energy scales $Q^2=100$~GeV$^2$ (top panel) and $m_Z^2$ (lower panel).
	}
\label{fig:fig5}
\end{figure}

In Fig.~\ref{fig:fig5}, to study the evolution effects  on the total strange distribution of the proton, 
we have made a similar comparison as in Fig.~\ref{fig:fig4}, but
for higher energy scales $Q^2=100$~GeV$^2$ (upper panel) and $Q^2=m_Z^2=8317$~GeV$^2$ (lower panel).
As is seen, in comparison with the initial energy scale (Fig.~\ref{fig:fig4}), a similar behavior is regained for the total strange distribution of  proton, but with a smaller intensity. It should be also noted that the inclusion of an IS component into the analysis leads to a  small change in the free parameter $f_s$ which appears in $ A_{\bar{s}}=f_s A_{\bar{d}}/(1-f_s)$.
For example, considering the second scenario the value of $f_s$ is changed from $0.7166$ (corresponding to the
analysis without considering an IS component) to $0.7323$ (for the analysis including an IS component with the probability of $2.5\%$).
The inclusion effect of an IS component can also lead to a small change in the contribution of total strange and the distribution of down-like sea quark.
Basically, when the IS probability ${\cal P}_5^{s\bar{s}}$ is increased in the analysis, a larger value is obtained for the $f_s$. 

In the following, we compute  the ratio of strange-to-light sea-quark densities in the proton for two situations: with and  without inclusion of
an IS component.
Analytically, the global PDF fits including the dimuon production 
in neutrino scattering \cite{Mason:2007zz,Samoylov:2013xoa,KayisTopaksu:2011mx,Goncharov:2001qe} suggest the strangeness suppression in the proton \cite{Dulat:2015mca,Harland-Lang:2014zoa,Alekhin:2014sya}. 
However, in Ref.~\cite{Aad:2012sb} it is shown that
the QCD analysis of the ATLAS 2010 $W^\pm$ and $Z$ measurements \cite{Aad:2011dm} offer the presence of strange quark at $x=0.023$ and $Q^2=1.9$~GeV$^2$.  In other words, their results assert the existence of all light quarks
in a similar strength in the kinematic range mentioned. This claim has been confirmed by the analysis of
ATLAS measurements of associated $W^\pm$ and charm production \cite{Aad:2014xca} and also the recent ATLAS 2011 $W^\pm$ and $Z$ measurements \cite{Aaboud:2016btc}.

To evaluate the fraction of the strange-quark density in the proton, we apply the following definition of the quantity $ r_s $ \cite{Aaboud:2016btc}
\begin{eqnarray}\label{eq9}
r_s=\frac{s+\bar s}{2\bar d},
\end{eqnarray}
which is the ratio of the strange to the down sea-quark distributions for a specific value of Bjorken variable $x$.
The analysis of Refs.~\cite{Aad:2014xca} and~\cite{Aaboud:2016btc} which are based on the ATLAS 2010 
and 2011 $W^\pm/Z$ measurements have determined the values $r_s=1.00$ and $r_s=1.19$, respectively,
at $x=0.023$ and $Q^2=1.9$~GeV$^2$. Since in our analysis, both the ATLAS 2010 and 2011 data sets
have been included simultaneously, then it would be interesting if we calculate the $ r_s $ for the analysis 
without considering an IS component.
For the extrinsic analysis, we obtain the value $r_s=1.18_{-0.06}^{+0.066} $ at $x=0.023$ and $Q^2=1.9$~ GeV$^2$. 
This value is a little smaller than the one obtained in the analysis of ATLAS~\cite{Aaboud:2016btc} Collaboration.
Note that, to determine this number they have just used their measurements along with the HERA I+II
combined data~\cite{Abramowicz:2015mha}, while our analysis includes a wide range of the collider data.
If we include an IS component in the analysis, the values of $r_s=1.20_{-0.061}^{+0.066}$ and $r_s=1.24_{-0.062}^{+0.067}$ are obtained considering the ${\cal P}_5^{s\bar{s}}=0.01$ and  ${\cal P}_5^{s\bar{s}}=0.025$, respectively, in the same kinematical range.
This means, the inclusion of an IS component in the
proton sea leads to a larger strange-quark density. In fact, when the probability of IS
distribution is increased a larger value is
obtained for the ratio $r_s$ in the kinematical range $x=0.023$ and $Q^2=1.9$~GeV$^2$.

\section{Summary and Conclusions}\label{sec:six}
For many years, various attempts have been made to study the impact of an intrinsic charm component 
of the nucleon sea on the global analysis of parton distribution functions~\cite{Harris:1995jx,Pumplin:2007wg,Nadolsky:2008zw,Dulat:2013hea,Jimenez-Delgado:2014zga}.
In this regard, some studies have been done to extract the intrinsic light quark distributions and the corresponding probabilities in the nucleon~\cite{Chang:2011vx,Chang:2011du,Chang:2014lea,Salajegheh:2015xoa,An:2017flb}. Although, a simultaneous study of the intrinsic and  extrinsic flavors  have not been performed yet in any global analysis of PDFs.
In the present work, we studied for the first time this simultaneous analysis in the case of intrinsic strange component
by virtue of the xFitter program~\cite{Alekhin:2014irh} and considering a wide range of the experimental data
such as the HERA I+II combined data~\cite{Abramowicz:2015mha} and the ATLAS $ W^\pm$, $Z$ 2010~\cite{Aad:2011dm} and 2011~\cite{Aaboud:2016btc} measurements. \\
For our aim, we applied the analytical result of BHPS model~\cite{Brodsky:1980pb,Brodsky:1981se} for the IS distribution, as the intrinsic contribution to the total strange density of the proton. Considering the BHPS model we employed two scenarios: in the first scenario we used the result of this model directly, and in the second one we used its evolved IS distribution. The result of BHPS model  leads to a momentum fraction $\langle x \rangle_{\textrm{IS}}=0.41$  while its evolved leads to a smaller one as $\langle x \rangle_{\textrm{IS}}=0.27$. 
As a primary conclusion we found that for both scenarios most of data sets have either a little sensitivity or no sensitivity on the IS component. This means, their $\chi^2$-value dose not have a 
considerable deviation from the one obtained through our basic analysis in the absence of IS component.
The most  variations for the $ \chi^2 $ arise from the NC HERA I+II 
combined data, the CMS inclusive jets~\cite{Chatrchyan:2012bja}
and the BCDMS proton structure functions $F_2^p$ \cite{Benvenuti:1989rh}.
Among all data sets, the Z rapidity distributions at 7~TeV from
the ATLAS 2010 data set and its 2011 data set for the forward channel  in the Z-peak regions
show a reduction in the $ \chi^2 $ when an IS component is considered in the proton sea.
Moreover, we found that if one follows the second scenario the sensitivity of experimental data on the IS component is less.\\
For each scenario, we also presented the limit of 
the IS probability $ {\cal P}_5^{s\bar{s}} $ for both standard tolerance criteria $ \Delta\chi^2=1 $ and 
$ 18.112 $ corresponding to $ 16 $ free parameters of the global fit, and also at the $ 1\sigma $ and $ 4\sigma $ level.
Following the first scenario we found that with $ \Delta\chi^2=1 $ at the $ 1\sigma $ level, the experimental
data do not tolerate any IS component in the proton. However, at the $ 4\sigma $ level 
an IS component with  the probability $ {\cal P}_5^{s\bar{s}}\approx 0.007 $ can be considered.
In analogy, for $ \Delta\chi^2=18.112 $ we found the $ {\cal P}_5^{s\bar{s}}\approx 0.008 $
and $ {\cal P}_5^{s\bar{s}}\approx 0.016 $ at the $1\sigma$ and $ 4\sigma $ level, respectively.
We have also concluded that following the second scenario and using the evolved BHPS IS distribution, the experimental data can tolerate an IS component 
with a greater probability $ {\cal P}_5^{s\bar{s}} $. In fact, with $ \Delta\chi^2=1 $ one gets $ {\cal P}_5^{s\bar{s}}\approx 0.001 $ and $ {\cal P}_5^{s\bar{s}}\approx 0.01 $ at the $ 1\sigma $ and $ 4\sigma $ level, respectively. 
The corresponding values for $ \Delta\chi^2=18.112 $ are $ {\cal P}_5^{s\bar{s}}\approx 0.011 $ at the $ 1\sigma $ and
$ {\cal P}_5^{s\bar{s}}\approx 0.025 $ at the $ 4\sigma $ level.\\
As a next step, we studied the strange content of the proton in detail. In this respect, using the second scenario we compared the proton extrinsic strange distribution (obtained through the global QCD analysis by ignoring the IS component)  with the results of the total strange distribution
obtained by including an IS component with the probabilities $1\%$ and $2.5\%$.
We found that the main changes in the total strange distribution of the proton sea (due to the inclusion of
an IS component) can be divided to two different regions: (1) an enhancement in the region $ 0 \lesssim x \lesssim 0.5 $ 
and (2) a reduction for $ x\gtrsim 0.5 $. We have shown that, as the probability of the IS distribution increases the rate of variations would be more.\\
In this work, the effect of evolution on the total strange distribution of the proton is also studied. 
It is shown that a similar behavior is obtained at higher energies.
The only remarkable point is that, as $ Q^2 $ increases the rate  of  enhancement and reduction  decreases, partially.\\
As a last conclusion, we calculated  the ratio of strange-to-light sea-quark densities $ r_s $ in the proton both by including and excluding  an IS component. For the extrinsic analysis, we obtained $r_s=1.18_{-0.06}^{+0.066}$ 
at $x=0.023$ and $Q^2=1.9$~GeV$^2$ which  is a little smaller than the result obtained in the 
recent analysis of ATLAS 2011 $W^\pm$ and $Z$ measurements~\cite{Aaboud:2016btc}.
We have shown that if one includes an IS component in the analysis, the values  $r_s=1.20_{-0.061}^{+0.066}$ and $r_s=1.24_{-0.062}^{+0.067}$ are obtained for
the analysis with the ${\cal P}_5^{s\bar{s}}=0.01$ and  ${\cal P}_5^{s\bar{s}}=0.025$, respectively.
Indeed, as the probability of IS distribution increases, a larger value is obtained for the
$r_s$ in the kinematical range $x=0.023$ and $Q^2=1.9$~GeV$^2$, as is expected.

\section*{Appendix: The extracted PDF sets}

For further studies on the impact of IS distributions on
the physical observables, it is important to have the PDF sets including the contribution of IS component.
Then, we have provided the results of our global PDF analysis  in the LHAPDF6 format,
including the basic analysis (where the contribution of an IS component is ignored) and the analysis of scenario II considering an IS component with the probabilities $1\%$ and $2.5\%$.
The grid files can be obtained via email from the authors. 

%

\end{document}